\documentclass[english]{article}
\usepackage[T1]{fontenc}
\usepackage[latin9]{inputenc}
\usepackage{bm}
\usepackage{amsmath}
\usepackage{amsthm}
\usepackage{amssymb}
\usepackage{stmaryrd}
\usepackage{graphicx}
\usepackage[authoryear]{natbib}

\makeatletter

\providecommand{\tabularnewline}{\\}

\newcommand{\lyxaddress}[1]{
	\par {\raggedright #1
	\vspace{1.4em}
	\noindent\par}
}
\theoremstyle{plain}
\newtheorem{prop}{\protect\propositionname}
\theoremstyle{remark}
\newtheorem{rem}{\protect\remarkname}

\usepackage{url}

\makeatother

\usepackage{babel}
\providecommand{\propositionname}{Proposition}
\providecommand{\remarkname}{Remark}

\begin{document}
\title{A binary-response regression model based on support vector machines}
\author{Hien D Nguyen$^{1}$ and Daniel V Fryer$^{2}$}
\maketitle

\lyxaddress{$^{1}$Department of Mathematics and Statistics, La Trobe University,
Bundoora, Victoria Australia. (Email: h.nguyen5@latrobe.edu.au) $^{2}$School
of Mathematics and Physics, University of Queensland, St Lucia, Queensland
Australia}
\begin{abstract}
The soft-margin support vector machine (SVM) is a ubiquitous tool
for prediction of binary-response data. However, the SVM is characterized
entirely via a numerical optimization problem, rather than a probability
model, and thus does not directly generate probabilistic inferential
statements as outputs. We consider a probabilistic regression model
for binary-response data that is based on the optimization problem
that characterizes the SVM. Under weak regularity assumptions, we
prove that the maximum likelihood estimate (MLE) of our model exists,
and that it is consistent and asymptotically normal. We further assess
the performance of our model via simulation studies, and demonstrate
its use in real data applications regarding spam detection and well
water access.
\end{abstract}

\section{Introduction}

Let $Y\in\left\{ -1,1\right\} $ be a binary response and let $\bm{X}\in\mathbb{X}\subseteq\mathbb{R}^{d}$
be some covariates. Furthermore, let $\mathcal{Z}_{n}=\left\{ \bm{Z}_{i}\right\} _{i=1}^{n}$,
where $\bm{Z}_{i}^{\top}=\left(\bm{X}_{i}^{\top},Y_{i}\right)$, be
an independent and identically distributed (IID) random sample of
$n\in\mathbb{N}$ pairs of response and covariates.

A common problem that arises when considering binary response variables
is to use the data $\mathcal{Z}_{n}$ to construct some discriminant
function $g:\mathbb{X}\rightarrow\left\{ -1,1\right\} $, such that
the probability of misclassification: $\Pr\left(g\left(\bm{X}\right)\ne Y\right)$,
is small (cf. \citealp[Ch. 1]{devroye1996probabilistic}). In \citet{cortes1995support},
the authors proposed the so-called (linear soft-margin) support vector
machine (SVM), whereupon the function $g$ was proposed to take the
form

\begin{equation}
g\left(\bm{x}\right)=\text{sign}\left(\alpha+\bm{x}^{\top}\bm{\beta}\right)\text{,}\label{eq: SVM rule}
\end{equation}
where $\bm{\theta}^{\top}=\left(\alpha,\bm{\beta}^{\top}\right)\in\mathbb{T}\subseteq\mathbb{R}^{d+1}$
are the parameters of $g$, and $\text{sign\ensuremath{\left(x\right)}}$
is equal to 1 if $x\ge0$, and equal to 0, otherwise. Here, we will
say that $\alpha$ is the intercept term and $\bm{\beta}^{\top}=\left(\beta_{1},\dots,\beta_{d}\right)$
is a vector of coefficients, where $\beta_{j}$ is the coefficient
of covariate $j\in\left[d\right]=\left\{ 1,\dots,d\right\} $. In
order to estimate the parameters $\bm{\theta}$ from $\mathcal{Z}_{n}$,
\citet{cortes1995support} suggested an optimization process that
is equivalent to solving the following problem (cf. \citealp{shawe2011review}):

\begin{equation}
\underset{\bm{\theta}\in\mathbb{T}}{\arg\min}\text{ }\frac{1}{n}\sum_{i=1}^{n}l\left(\bm{Z}_{i};\bm{\theta}\right)+\lambda\bm{\beta}^{\top}\bm{\beta}\text{,}\label{eq: svm optim}
\end{equation}
where $\lambda>0$ is a regularization constant for the size of $\bm{\beta}$,
and

\[
l\left(\bm{Z}_{i};\bm{\theta}\right)=\left[1-Y_{i}\tilde{\bm{X}}_{i}^{\top}\bm{\theta}\right]_{+}
\]
is a loss function with $\tilde{\bm{x}}^{\top}=\left(1,\bm{x}^{\top}\right)$
and $\left[x\right]_{+}=\max\left\{ x,0\right\} $. The SVM has become
a ubiquitously successful tool for data analysts and applied researchers,
and its virtues are well-exposed in volumes such as \citet{abe2005support},
\citet{Chen:2004aa}, \citet{liang2016support}, and \citet{murty2016support}.

Noting the form of the optimization problem, \citet{polson2011data}
proposed that one can consider the equivalent optimization routine
(for $\lambda=0$)

\begin{equation}
\underset{\bm{\theta}\in\mathbb{T}}{\arg\max}\text{ }-\sum_{i=1}^{n}l\left(\bm{Z}_{i};\bm{\theta}\right)\label{eq: approximate MLE}
\end{equation}
to be an approximation of the maximum likelihood estimation (MLE)
of $\bm{\theta}$ under the probability model:
\begin{align}
\Pr\left(Y_{i}=y_{i}|\bm{X}_{i}=\bm{x}_{i}\right) & =f\left(y_{i}|\bm{x}_{i};\bm{\theta}\right)\nonumber \\
 & =\frac{\exp\left(-\left[1-y_{i}\tilde{\bm{x}}_{i}^{\top}\bm{\theta}\right]_{+}\right)}{\exp\left(-\left[1-\tilde{\bm{x}}_{i}^{\top}\bm{\theta}\right]_{+}\right)+\exp\left(-\left[1+\tilde{\bm{x}}_{i}^{\top}\bm{\theta}\right]_{+}\right)}\text{,}\label{eq: svmit}
\end{align}
whereby the normalization term in the denominator of (\ref{eq: svmit})
is omitted. This approximation was also used by \citet{fu2010mixing},
\citet{mao2014nonlinear}, \citet{lai2015mixture}, and \citet{wenzel2017bayesian},
where it is argued that it is more computationally feasible than MLE
(since it is a concave optimization problem) and that the functional
form more closely resembles the SVM problem from which it is derived. 

In this paper we consider the MLE problem of computing
\begin{equation}
\hat{\bm{\theta}}_{n}=\underset{\bm{\theta}\in\mathbb{T}}{\arg\max}\text{ }l_{n}\left(\bm{\theta}\right)\text{,}\label{eq: MLE}
\end{equation}
where
\begin{equation}
l_{n}\left(\bm{\theta}\right)=n^{-1}\sum_{i=1}^{n}\log f\left(Y_{i}|\bm{X}_{i};\bm{\theta}\right)\text{,}\label{eq: log-likelihood}
\end{equation}
instead of (\ref{eq: approximate MLE}). We prove that the log-likelihood
function (\ref{eq: log-likelihood}) is coercive, on average, conditional
on the covariates $\left\{ \bm{X}_{i}\right\} _{i=1}^{n}$, thus guaranteeing
the existence of a global maximizer of the limiting function within
the interior of some compact subset of $\mathbb{T}=\mathbb{R}^{d+1}$.
This is sufficient for establishing consistency of the estimator.
Furthermore, recent evidence suggests that one can compute the maximum
of non-convex and non-differentiable functions, such as (\ref{eq: log-likelihood}),
using quasi-Newton methods such as the Broyden--Fletcher--Goldfarb--Shanno
(BFGS; \citealp{Fletcher1987}) algorithm (see, e.g., \citealp{lewis2013nonsmooth}
and \citealp{keskar2019limited}), especially with the aid of automatic
differentiation (AD; see, e.g., \citealp{bucker2006automatic}). Regarding
the maximum likelihood estimator (MLE), we further demonstrate that
one can establish conditions under which consistence and asymptotic
normality hold, and thus permit drawing of inference via model (\ref{eq: svmit}).

We assess the performance of our approach via a finite sample assessment
of its asymptotic properties in simulation studies. Here, we also
assess how well the model performs prediction of an unknown response
$Y$ given some observed covariate $\bm{x}$ in a similar manner to
an SVM and logistic regression (see, e.g., \citealp[Ch. 8]{McLachlan:1992aa},
and \citealp{hosmer2013applied}). We then apply our method to a pair
of real-world data sets, regarding spam detection and well water access,
and compare the inference drawn from model (\ref{eq: svmit}) to those
drawn via logistic regression, as well as its ability to conduct prediction
as compared to a SVM.

The paper proceeds as follows. In Section 2, we consider the existence,
consistency and asymptotic normality of the MLE. In Section 3, we
describe our computation strategy and conduct simulation studies.
In Section 4, we present example applications. Finally, we present
some concluding remarks in Section 5.

\section{The maximum likelihood estimator}

\subsection{\label{subsec:Existence}Existence}

In order for maximum likelihood estimation (MLE) to make sense, we
must demonstrate that the MLE exists in some useful sense. To that
effect, we wish to show that conditional on $\left\{ \bm{X}_{i}\right\} _{i=1}^{n}=\left\{ \bm{x}_{i}\right\} _{i=1}^{n}$
(for brevity, we shall write $\mathcal{X}_{n}=\left\{ \bm{X}_{i}\right\} _{i=1}^{n}$
and $\mathbf{x}_{n}=\left\{ \bm{x}_{i}\right\} _{i=1}^{n}$), the
expected value of the log-likelihood (\ref{eq: log-likelihood}) has
all of its global maxima, with respect to $\bm{\theta}\in\mathbb{T}=\mathbb{R}^{d}$,
in the interior $\text{int}\left(\mathbb{S}\right)$ of some compact
set $\mathbb{S}\subset\mathbb{T}$, for each $n$. This can be achieved
by showing that $-\mathbb{E}\left[l_{n}\left(\bm{\theta}\right)|\mathcal{X}_{n}=\mathbf{x}_{n}\right]$
is coercive, in the sense that
\begin{equation}
-\mathbb{E}\left[l_{n}\left(\bm{\theta}\right)|\mathcal{X}_{n}=\mathbf{x}_{n}\right]\rightarrow\infty\text{, if }\left\Vert \bm{\theta}\right\Vert \rightarrow\infty\text{,}\label{eq: expected coerciveness}
\end{equation}
for each $i\in\left[n\right]$ (cf. \citealp[Sec. 3.1]{Auslender:2002aa}),
where $\left\Vert \cdot\right\Vert $ denotes the Euclidean norm.

Let $p_{i}=f\left(1|\bm{x}_{i};\bm{\theta}_{0}\right)\in\left(0,1\right)$
and let $\tilde{p}_{i}=1-p_{i}$. Here, $\bm{\theta}_{0}\in\mathbb{T}$
is the true value of $\bm{\theta}$, which arises from the data generating
process of $\mathcal{Z}_{n}$. Then, we may write
\begin{eqnarray*}
-\mathbb{E}\left[\log f\left(1|\bm{x}_{i};\bm{\theta}\right)\right] & = & p_{i}\left[1-\tilde{\bm{x}}_{i}^{\top}\bm{\theta}\right]_{+}+\tilde{p}_{i}\left[1+\tilde{\bm{x}}_{i}^{\top}\bm{\theta}\right]_{+}\\
 &  & +\log\left[\exp\left(-\left[1-\tilde{\bm{x}}_{i}^{\top}\bm{\theta}\right]_{+}\right)+\exp\left(-\left[1+\tilde{\bm{x}}_{i}^{\top}\bm{\theta}\right]_{+}\right)\right]\text{,}
\end{eqnarray*}
for each $i\in\left[n\right]$.

\begin{figure}
\begin{centering}
\includegraphics[width=10cm]{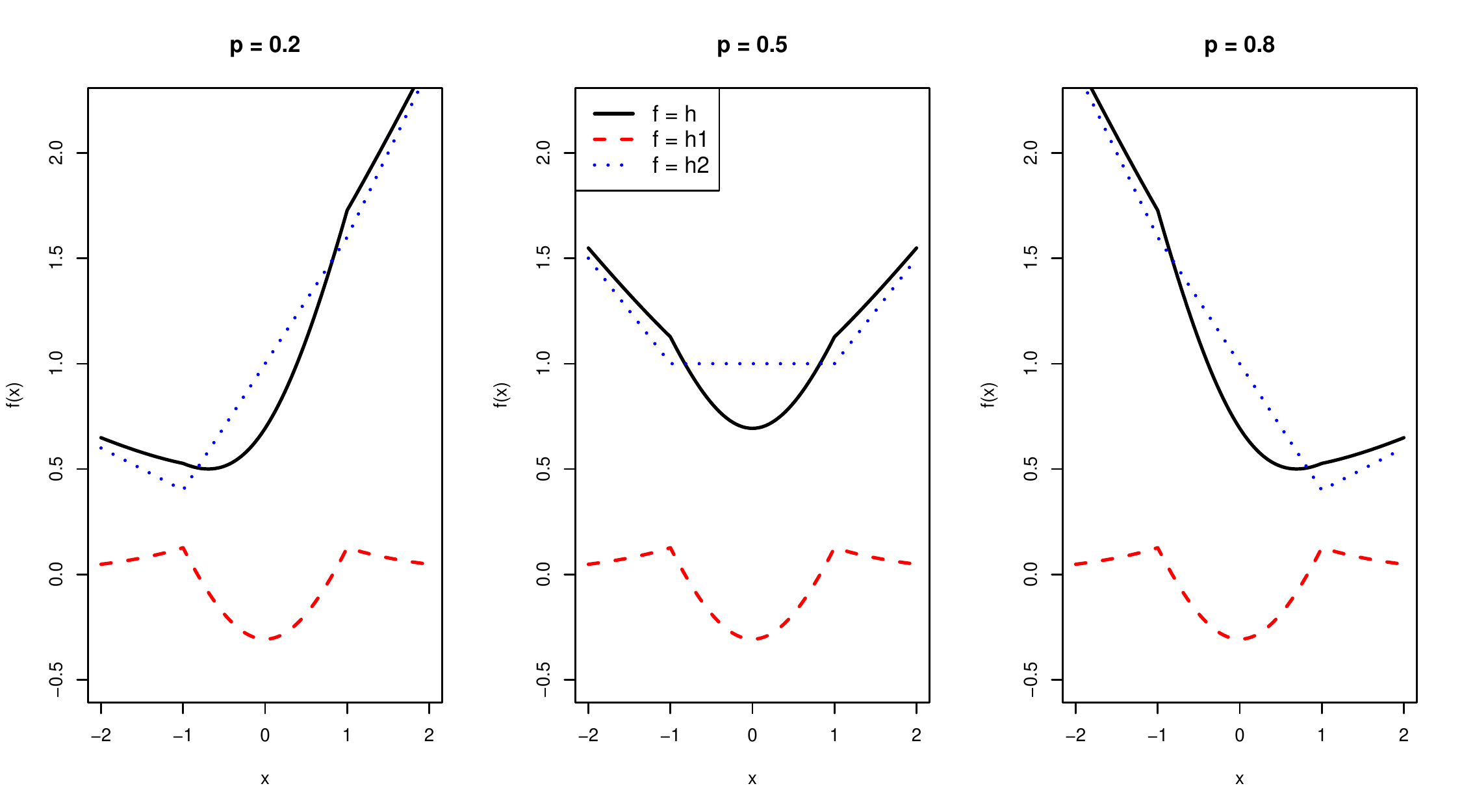}
\par\end{centering}
\centering{}\caption{\label{fig: h1h2}Visualization of the components of (\ref{eq: expected log like}),
for $p_{i}\in\left\{ 0.2,0.5,0.8\right\} $.}
\end{figure}

Consider the substitution $\tilde{\bm{x}}_{i}^{\top}\bm{\theta}=\tilde{\theta}_{i}$
and thus write
\begin{equation}
-\mathbb{E}\left[\log f\left(1|\bm{x}_{i};\bm{\theta}\right)\right]=h\left(\tilde{\theta}_{i}\right)=h_{1}\left(\tilde{\theta}_{i}\right)+h_{2}\left(\tilde{\theta}_{i}\right)\text{,}\label{eq: expected log like}
\end{equation}
where
\[
h_{1}\left(\tilde{\theta}_{i}\right)=p_{i}\left[1-\tilde{\theta}_{i}\right]_{+}+\tilde{p}_{i}\left[1+\tilde{\theta}_{i}\right]_{+}\text{,}
\]
and
\[
h_{2}\left(\tilde{\theta}_{i}\right)=\log\left[\exp\left(-\left[1-\tilde{\theta}_{i}\right]_{+}\right)+\exp\left(-\left[1+\tilde{\theta}_{i}\right]_{+}\right)\right]\text{.}
\]
We firstly wish to show that (\ref{eq: expected log like}) is coercive,
with respect to $\tilde{\theta}_{i}$. A visualization of $h$, $h_{1}$
and $h_{2}$ appears in Figure \ref{fig: h1h2}.

We may inspect $h_{2}$ at its limits and local extrema and observe
that $\left|h_{2}\left(\tilde{\theta}_{i}\right)\right|\le1-\log\left(2\right)$,
for all $\tilde{\theta}_{i}$. Thus, $h_{2}$ is a bounded function.
We observe that $h_{1}$ is coercive since $h_{1}\left(\tilde{\theta}_{i}\right)\rightarrow\infty$
for $\tilde{\theta}_{i}\rightarrow\pm\infty$. We thus also establish
that $h=h_{1}+h_{2}$ is coercive since $h_{2}$ is a function that
is bounded from below (and in this case, also above). 

Next, we appeal to Lemma 5.1 of \citet{calatroni2019flexible} (see
also \citealp[Lem. 2.7.1]{ciak2015coercive}), which implies that
if $h\left(\theta_{i}\right)$ is a proper, continuous, and coercive
function, and if we have the null space condition:

\[
\text{null}\left(\tilde{\bm{x}}_{i}\right)=\left\{ \bm{\theta}\in\mathbb{R}^{d}:\tilde{\bm{x}}_{i}^{\top}\bm{\theta}=0\right\} =\left\{ \mathbf{0}\right\} \text{,}
\]
then (\ref{eq: expected log like}) is continuous and coercive, with
respect to $\bm{\theta}$. Here $\mathbf{0}$ denotes the zero vector,
and proper is taken to mean that $h\left(\tilde{\theta}_{i}\right)<\infty$
for at least one $\theta_{i}$ and $h\left(\tilde{\theta}_{i}\right)>-\infty$
for all $\tilde{\theta}_{i}\in\mathbb{R}$. Since $h$ is univariate,
coercive, and bounded below, we automatically have the fact that $h$
is proper, and thus (\ref{eq: expected log like}) is coercive as
long as $\text{null}\left(\tilde{\bm{x}}_{i}\right)=\left\{ \bm{0}\right\} $.
Since the sum of coercive functions is coercive, we obtain the following
result.
\begin{prop}
\label{prop: coercivity}If $\bigcap_{i\in\left[n\right]}\mathrm{null}\left(\tilde{\bm{x}}_{i}\right)=\left\{ \mathbf{0}\right\} $,
then the expected conditional log-likelihood $\mathbb{E}\left[l_{n}\left(\bm{\theta}\right)|\mathcal{X}_{n}=\mathbf{x}_{n}\right]$
is coercive, and thus there exists some compact set $\mathbb{S}\subset\mathbb{T}=\mathbb{R}^{d+1}$,
such that the set of global maxima
\[
\underset{\bm{\theta}\in\mathbb{T}}{\arg\max}\text{ }\mathbb{E}\left[l_{n}\left(\bm{\theta}\right)|\mathcal{X}_{n}=\mathbf{x}_{n}\right]
\]
is equal to
\[
\underset{\bm{\theta}\in\mathrm{int}\left(\mathbb{S}\right)}{\arg\max}\text{ }\mathbb{E}\left[l_{n}\left(\bm{\theta}\right)|\mathcal{X}_{n}=\mathbf{x}_{n}\right]\text{.}
\]
\end{prop}
From Proposition \ref{prop: coercivity}, we may conclude that (\ref{eq: MLE})
exists, so long we do not observe data $\mathcal{Z}_{n}$, where all
of the responses are equal to $-1$ or are all equal to $1$, and
where we do not observe some pathological set of covariates $\mathcal{X}_{n}$,
where $\bigcap_{i\in\left[n\right]}\text{null}\left(\tilde{\bm{X}}_{i}\right)\ne\left\{ \mathbf{0}\right\} $.
One potential pathology is if all of the vectors of $\mathcal{X}_{n}$
are linearly dependent. For $\bm{X}_{i}$ arising from some continuous
distribution, this event will occur with probability zero.

\subsection{Consistency}

We begin by establishing the consistency of the MLE over some arbitrarily
large compact subset $\mathbb{S}$ of $\mathbb{T}$, as we are permitted
to do via Proposition \ref{prop: coercivity}. Further assume that
$\mathbb{X}$ is a compact subset of $\mathbb{R}^{d}$. To this end,
we firstly consider the limit of (\ref{eq: log-likelihood}) conditional
on $\mathcal{X}_{n}=\mathbf{x}_{n}$, for fixed $\bm{\theta}\in\mathbb{T}$.
Using the independent but not identical law of large numbers of \citet[Cor. 3.9]{White:2001aa},
we have the fact that
\begin{equation}
l_{n}\left(\bm{\theta}\right)\overset{\text{p}}{\longrightarrow}\mathbb{E}\left[l_{n}\left(\bm{\theta}\right)|\mathcal{X}_{n}=\mathbf{x}_{n}\right]\text{,}\label{eq: lln}
\end{equation}
as $n\rightarrow\infty$, conditional on the existence of some constant
$C<\infty$, such that $\mathbb{E}\left[\log^{2}f\left(Y_{i}|\bm{x}_{i};\bm{\theta}\right)\right]<C$,
for all $i$, and fixed $\bm{\theta}$. Here $\overset{\text{p}}{\longrightarrow}$
denotes convergence in probability. This is easy to verify, since
$Y_{i}\in\left\{ -1,1\right\} $ is a discrete random variable and
$\mathbb{X}$ is compact, so we may take
\[
C=\sup_{\bm{x}\in\mathbb{X}}\:\left[\log^{2}f\left(-1|\bm{x};\bm{\theta}\right)+\log^{2}f\left(1|\bm{x};\bm{\theta}\right)\right]\text{.}
\]

Next, we must make the convergence in probability uniform over some
compact set $\mathbb{S}\subseteq\mathbb{T}$. That is, we require
that
\begin{equation}
\sup_{\bm{\theta}\in\mathbb{S}}\left|l_{n}\left(\bm{\theta}\right)-\mathbb{E}\left[l_{n}\left(\bm{\theta}\right)|\mathcal{X}_{n}=\mathbf{x}_{n}\right]\right|\overset{\text{p}}{\longrightarrow}0\text{.}\label{eq: ulln}
\end{equation}
We can verify this using the generic uniform law of large numbers
of \citet[Cor. 3.1]{newey1991uniform}. This can be established by
verifying that $\mathbb{S}$ is compact, that (\ref{eq: lln}) is
satisfied, and that the Lipschitz condition 
\begin{equation}
\left|\log f\left(y|\bm{x};\bm{\theta}\right)-\log f\left(y|\bm{x};\bm{\vartheta}\right)\right|\le L\left\Vert \bm{\theta}-\bm{\vartheta}\right\Vert \text{,}\label{eq: lipschitz}
\end{equation}
for each fixed $\left(y,\bm{x}\right)\in\left\{ -1,1\right\} \times\mathbb{X}$,
where $\bm{\theta},\bm{\vartheta}\in\mathbb{S}$ and $L<\infty$ is
a constant.

As in (\ref{eq: expected log like}), we consider the map $\tilde{\theta}=\tilde{\bm{x}}^{\top}\bm{\theta}$
. Let $\mathbb{I}\subset\mathbb{R}$ be a sufficiently large compact
interval such that $\tilde{\bm{x}}^{\top}\bm{\theta}\in\mathbb{I}$
for all values of $\bm{x}\in\mathbb{X}$ and $\bm{\theta}\in\mathbb{S}$.
This is possible via the compactness of $\mathbb{S}$ and $\mathbb{X}$.
Next we wish to establish the fact that 
\[
\tilde{h}\left(\theta\right)=-\left[1-y\tilde{\theta}\right]_{+}+\log\left[\exp\left(-\left[1-y\tilde{\theta}\right]_{+}\right)+\exp\left(-\left[1+y\tilde{\theta}\right]_{+}\right)\right]
\]
 s Lipschitz for any $y\in\left\{ -1,1\right\} $, with respect to
$\tilde{\theta}\in\mathbb{I}$. This can be achieved by noting that
$\tilde{h}$ is piecewise continuously differentiable, and by applying
\citet[Cor. 4.1.1]{scholtes2012introduction}. Using the affine map
$\tilde{\theta}=\tilde{\bm{x}}^{\top}\bm{\theta}$, from $\mathbb{S}$
to $\mathbb{I}$, we establish (\ref{eq: lipschitz}) by the fact
that Lipschitz compositions are Lipschitz. Thus, (\ref{eq: ulln})
is verified. 

By the continuity of (\ref{eq: log-likelihood}) and its uniform convergence
in probability (\ref{eq: ulln}), we can now apply \citet[Lem. 5]{nguyen2016laplace},
a non-smooth version of the extremum estimator consistency theorem
of \citet{amemiya1985advanced}, in order to establish the following
consistency result regarding (\ref{eq: MLE}).
\begin{prop}
\label{prop: consistency}Let $\mathbb{S}\subset\mathbb{T}$ and $\mathbb{X}\subset\mathbb{R}^{d}$
be compact, such that 
\[
\mathbb{S}_{n}=\left\{ \bm{\theta}\in\mathrm{int}\left(\mathbb{S}\right):\bm{\theta}\text{ is a local maximum of }l_{n}\left(\bm{\theta}\right)\right\} \text{,}
\]
and assume that $\mathbb{E}\left[l_{n}\left(\bm{\theta}\right)|\mathcal{X}_{n}=\mathbf{x}_{n}\right]$
attains a strict local maximum at $\bm{\theta}_{0}\in\mathrm{int}\left(\mathbb{S}\right)$.
Then, for any $\epsilon>0$, $\inf_{\bm{\theta}\in\mathbb{S}_{n}}\left\Vert \bm{\theta}-\bm{\theta}_{0}\right\Vert \overset{\mathrm{p}}{\longrightarrow}0$.
\end{prop}
This proposition is useful in the context of solving problem (\ref{eq: MLE})
since (\ref{eq: log-likelihood}) is likely to have multiple local
and global maxima, and similarly with the conditional expected log-likelihood
$\mathbb{E}\left[l_{n}\left(\bm{\theta}\right)|\mathcal{X}_{n}=\mathbf{x}_{n}\right]$.
The result ensures that if we follow the sequences of strict local
maxima of (\ref{eq: log-likelihood}), then we obtain sequences of
consistent estimators for each of the local maxima $\bm{\theta}_{0}$
of $\mathbb{E}\left[l_{n}\left(\bm{\theta}\right)|\mathcal{X}_{n}=\mathbf{x}_{n}\right]$.
Of course, in any one run of an optimization algorithm, one tends
to only find one local maximum. Thus, it is often advisable to run
the optimization algorithm for computing (\ref{eq: MLE}) multiple
times, with different initializations, in order to ensure that one
has located the local maximum that corresponds to the $\bm{\theta}_{0}$
one is seeking (cf. \citealp[Sec. 4.1.1]{amemiya1985advanced}).

\subsection{Asymptotic normality}

We may now establish asymptotic normality via \citet[Thm. 4.1.3]{amemiya1985advanced}.
Here, we require that the local maximum of interest $\bm{\theta}_{0}$
is such that the assumptions of Proposition \ref{prop: consistency}
are satisfied, and also that there is an open and convex neighborhood
around $\bm{\theta}_{0}$, where $\log f\left(y_{i}|\bm{x}_{i};\bm{\theta}\right)$
(for each $i\in\left[n\right]$) is three times differentiable with
respect to $\bm{\theta}\in\mathbb{S}$, and all first, second, and
third order partial derivatives are bounded (in order to apply \citealt[Thm. 4.1.4]{amemiya1985advanced}).
Here, the boundedness can be established simply via the fact that
$Y_{i}\in\left\{ -1,1\right\} $ is a discrete random variable. We
thus have the following result.
\begin{prop}
\label{prop: asymptotic norm}Assume that $\mathbb{S}\subset\mathbb{T}$
and $\mathbb{X}\subset\mathbb{R}^{d}$ are compact, and that $\log f\left(y|\bm{x};\bm{\theta}\right)$
is three times differentiable with respect to $\bm{\theta}$ in a
open and convex neighborhood of $\bm{\theta}_{0}$, for each $i\in\left[n\right]$.
Let $\left\{ \hat{\bm{\theta}}_{n}\right\} _{n=1}^{\infty}$ be a
sequence that is obtained by choosing one element of $\mathbb{S}_{n}$
(for each $n$), as defined in Proposition \ref{prop: consistency},
such that $\hat{\bm{\theta}}_{n}\overset{\text{p}}{\longrightarrow}\bm{\theta}_{0}$.
Then, $\sqrt{n}\left(\hat{\bm{\theta}}_{n}-\bm{\theta}_{0}\right)$
converges in law to a normal distribution with mean vector $\mathbf{0}$
and covariance matrix $\mathbf{A}^{-1}\left(\bm{\theta}_{0}\right)\mathbf{B}\left(\bm{\theta}_{0}\right)\mathbf{A}^{-1}\left(\bm{\theta}_{0}\right)$,
where $\mathbf{A}\left(\bm{\theta}_{0}\right)$ is assumed to be non-singular,
and
\end{prop}
\[
\mathbf{A}\left(\bm{\theta}\right)=\lim_{n\rightarrow\infty}\frac{1}{n}\sum_{i=1}^{n}\mathbb{E}\left[\left.\frac{\partial^{2}f\left(Y_{i}|\bm{X}_{i};\bm{\theta}\right)}{\partial\bm{\theta}\partial\bm{\theta}^{\top}}\right|_{\bm{\theta}}|\bm{X}_{i}=\bm{x}_{i}\right]
\]
 and 
\[
\mathbf{B}\left(\bm{\theta}\right)=\lim_{n\rightarrow\infty}\frac{1}{n}\sum_{i=1}^{n}\mathbb{E}\left[\left.\frac{\partial f\left(Y_{i}|\bm{X}_{i};\bm{\theta}\right)}{\partial\bm{\theta}}\right|_{\bm{\theta}}\left.\frac{\partial f\left(Y_{i}|\bm{X}_{i};\bm{\theta}\right)}{\partial\bm{\theta}^{\top}}\right|_{\bm{\theta}}|\bm{X}_{i}=\bm{x}_{i}\right]\text{.}
\]

We note that the covariance form $\mathbf{A}^{-1}\left(\bm{\theta}_{0}\right)\mathbf{B}\left(\bm{\theta}_{0}\right)\mathbf{A}^{-1}\left(\bm{\theta}_{0}\right)$
assumes that there may be misspecification between the model (\ref{eq: svmit})
and the data generating process of $\mathcal{Z}_{n}$ (cf. \citealp{white1982maximum}).
If there is no misspecification, then we may take $\mathbf{A}\left(\bm{\theta}_{0}\right)=-\mathbf{B}\left(\bm{\theta}_{0}\right)$
and thus the covariance matrix reduces to$-\mathbf{A}^{-1}\left(\bm{\theta}_{0}\right)=\mathbf{B}\left(\bm{\theta}_{0}\right)$.
We may estimate $\mathbf{A}\left(\bm{\theta}_{0}\right)$ and $\mathbf{B}\left(\bm{\theta}_{0}\right)$
by 
\begin{equation}
\hat{\mathbf{A}}_{n}\left(\hat{\bm{\theta}}_{n}\right)=\frac{1}{n}\sum_{i=1}^{n}\left.\frac{\partial^{2}\log f\left(Y_{i}|\bm{X}_{i};\bm{\theta}\right)}{\partial\bm{\theta}\partial\bm{\theta}^{\top}}\right|_{\hat{\bm{\theta}}_{n}}\label{eq: est A}
\end{equation}
and
\begin{equation}
\hat{\mathbf{B}}_{n}\left(\hat{\bm{\theta}}_{n}\right)=\frac{1}{n}\sum_{i=1}^{n}\left.\frac{\partial\log f\left(Y_{i}|\bm{X}_{i};\bm{\theta}\right)}{\partial\bm{\theta}}\right|_{\hat{\bm{\theta}}_{n}}\left.\frac{\partial\log f\left(Y_{i}|\bm{X}_{i};\bm{\theta}\right)}{\partial\bm{\theta}^{\top}}\right|_{\hat{\bm{\theta}}_{n}}\text{,}\label{eq: Est B}
\end{equation}
respectively, via the sample $\mathcal{Z}_{n}$ (see, e.g., \citealp[Thm. 7.3]{Boos:2013aa}).

\section{\label{sec:Implementation-and-numerical}Implementation and numerical
studies}

\subsection{Computational specifics}

We compute the MLE (\ref{eq: svmit}) using the BFGS method as implemented
in $\mathsf{R}$ \citep{R-Core-Team:2020aa} via the $\mathtt{optim}$
function. Here, we use gradients that are computed via AD using the
package $\mathsf{autodiffr}$ \citep{Li:2018aa}. It is established
in \citet{lewis2009nonsmooth}, \citet{lewis2013nonsmooth}, and \citet{keskar2019limited},
that the BFGS performs well in non-differentiable and non-convex settings.
In such situations, they prove that the line search steps are convergent
under general conditions, although it is difficult to prove the global
convergence of the algorithm, overall, except in the simple case of
the Euclidean norm function. However, via comprehensive simulation
studies, it is found that the BFGS method tends to be correct under
standard settings.

In order to guarantee global convergence, \citet{lewis2009nonsmooth}
and \citet{lewis2013nonsmooth} suggest that one should apply a gradient
sampling method after the BFGS solution is found. It is established
in \citet{burke2005robust} that gradient sampling is globally convergent
under standard settings, and we implemented the BFGS-then-gradient
(also referred to by \citealp{lewis2009nonsmooth} and \citealp{lewis2013nonsmooth}
as HANSO: hybrid algorithm for non-smooth optimization) sampling approach
via the $\mathsf{rHanso}$ package of \citet{Mallik:2013aa}. 

Via a battery of simulation settings, we found that the $\mathtt{hanso}$
function from $\mathsf{rHanso}$ produced exactly the same outcomes
as BFGS method using $\mathtt{optim}$ in many cases, and in other
cases was actually less optimal. Thus, since HANSO requires an gradient
sampling step, which is significantly more computationally intensive,
we opted to rely on the standard BFGS method only, for all of our
computations. Code, for some of the computation in this section and
the next, can be found online at: \url{https://github.com/hiendn/svm_binary_regression}.

\subsection{Finite sample accuracy of the MLE}

Although Proposition \ref{prop: consistency} implies that one can
always arbitrarily accurately estimate $\bm{\theta}_{0}$, that characterizes
the data generating process (\ref{eq: svmit}), with the MLE (\ref{eq: MLE})
using a sufficiently large IID sample $\mathcal{Z}_{n}$, it is unclear
as to how large $n$ needs to be in order for Proposition \ref{prop: consistency}
to apply. We thus assess the performance of the MLE when $n$ is a
finite value. Here we choose $n\in\left\{ 100,200,500,1000\right\} $. 

For each $n$, we simulate $\mathcal{Z}_{n}$ with each covariate
$\bm{X}_{i}$ arising from a multivariate normal distribution with
mean $\mathbf{0}$ and covariance matrix $\mathbf{I}$ (identity matrix),
for dimensions $d\in\left\{ 1,5,10\right\} $. We then simulate each
$Y_{i}$ using the model (\ref{eq: svmit}) with $\bm{\theta}_{0}^{\top}=\left(\alpha_{0},\bm{\beta}_{0}^{\top}\right)=\mathbf{1}^{\top}$
(the ones vector). An MLE $\hat{\bm{\theta}}_{n}$ is then computed
using the BFGS algorithm, as described above.

For each combination of $n$ and $d$, we repeat the simulation above
$R=100$ times. We then compute the mean squared error 
\[
\text{MSE}=\frac{1}{R}\sum_{r=1}^{R}\left\Vert \hat{\bm{\theta}}_{n}^{\left(r\right)}-\bm{\theta}_{0}\right\Vert ^{2}\text{,}
\]
 for each simulation combination and report it Table \ref{tab:Mean-squared-errors}.

\begin{table}
\caption{\label{tab:Mean-squared-errors}Mean squared errors from 100 replications
of the MLE (\ref{eq: MLE}) for various combinations of dimension
$d$ and sample size $n$. Here, $a\left(b\right)=a\times10^{b}$.}

\centering{}%
\begin{tabular}{|c|ccc|}
\hline 
 &  & $d=$ & \tabularnewline
$n=$ & 1 & 5 & 10\tabularnewline
\hline 
100 & 1.80(--1) & 2.75(+0) & 2.30(+3)\tabularnewline
200 & 7.27(--2) & 3.78(--1) & 3.48(+0)\tabularnewline
500 & 2.68(--2) & 1.57(--1) & 3.30(--1)\tabularnewline
1000 & 1.21(--2) & 6.12(--2) & 1.37(--1)\tabularnewline
2000 & 6.82(--3) & 2.85(--2) & 8.76(--2)\tabularnewline
\hline 
\end{tabular}
\end{table}

From Table \ref{tab:Mean-squared-errors} we observe, as expected,
that for fixed $n$, an increase in the dimensionality $d$ increases
the MSE, since there are more parameters to estimate and thus the
complexity of the problem increases. Furthermore, for fixed $d$,
we observe that as $n$ increases, the MSE decreases. This conforms
with with the conclusions from the consistency result of Proposition
\ref{prop: consistency}. We observe for the larger values of $n$
(500, 1000, and 2000), that the rate of decrease of the MSE is approximately
linear in $n$, which is as predicted by the asymptotic normality
result of Proposition \ref{prop: asymptotic norm}.

\begin{figure}
\begin{centering}
\includegraphics[width=10cm]{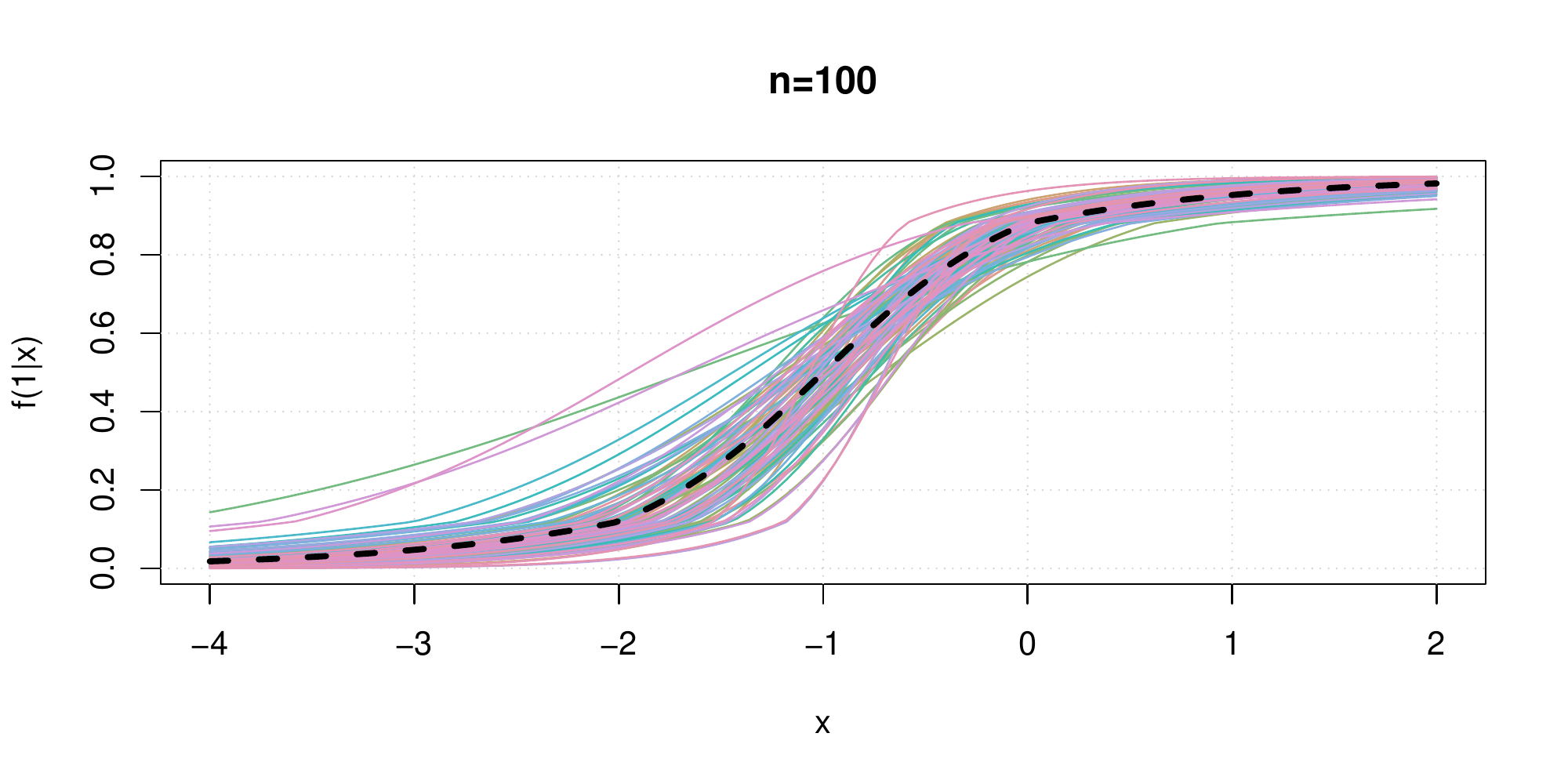}
\par\end{centering}
\begin{centering}
\includegraphics[width=10cm]{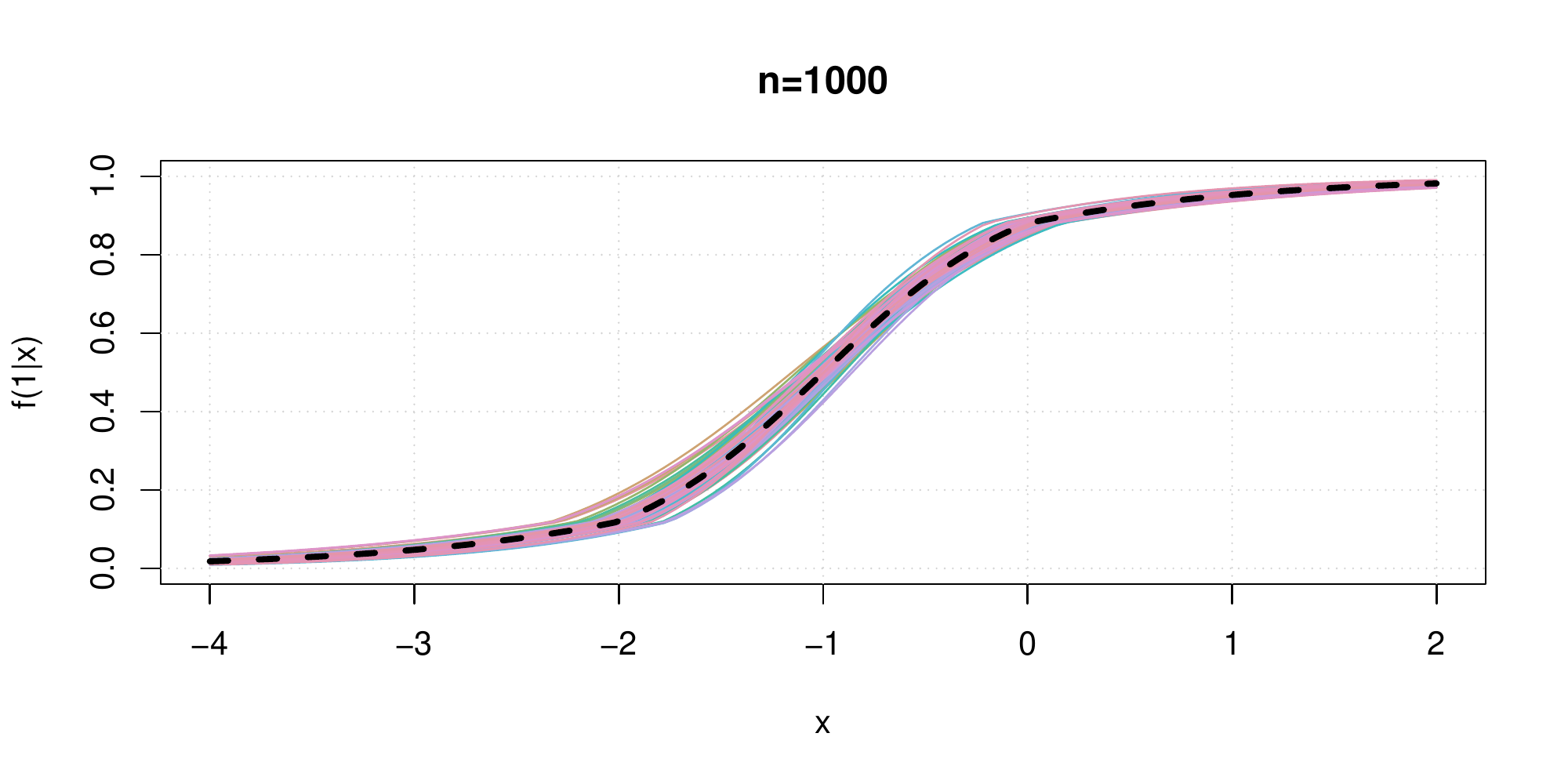}
\par\end{centering}
\begin{centering}
\caption{\label{fig: fitted curves}Curve of the estimated conditional probability
function $f\left(1|\bm{x},\hat{\bm{\theta}}_{n}^{\left(r\right)}\right)$
for each replication from the accuracy simulation for the $d=100$
scenario, where the MLE is computed using samples of sizes $n=100$
and $n=1000$. The dashed curve indicates the generative conditional
probability function, where $\bm{\theta}_{0}=\mathbf{1}$.}
\par\end{centering}
\end{figure}

In Figure \ref{fig: fitted curves} we plot the conditional probability
curves $f\left(1|\bm{x},\hat{\bm{\theta}}_{n}^{\left(r\right)}\right)$,
with respect to $\bm{x}$, corresponding to each of the replications
in the case of $d=1$, for $n=100$ and $n=1000$. We observe that
there is a dramatic increase in accuracy of the estimation of the
generative conditional probability curve, when $n$ is increased from
100 to 1000.

\subsection{Binary prediction accuracy}

Here we assess the ability of model (\ref{eq: svmit}), fitted via
MLE, to predict the value of $Y^{\prime}$ given some observed covariate
$\bm{X}^{\prime}$. This prediction is conducted in the same manner
as when performing prediction using logistic regression. That is,
we use the maximum a posteriori approach, whereupon we predict $Y$
via the rule:

\begin{equation}
\hat{y}\left(\bm{X}\right)=\arg\max_{y\in\left\{ -1,1\right\} }\text{ }\text{ }f\left(y|\bm{X};\hat{\bm{\theta}}_{n}\right)\text{.}\label{eq: MAP}
\end{equation}

In order to assess the performance of rule (\ref{eq: MAP}), we conduct
the following simulation study. A sample $\mathcal{Z}_{n}$ of $n\in\left\{ 100,1000\right\} $
pairs of responses and covariates are simulated from a two-component
normal mixture model (see, e.g., \citealp[Ch. 3]{McLachlan:2000aa}),
where each $\bm{X}_{i}$ is of dimension $d\in\left\{ 2,5\right\} $,
using the $\mathsf{MixSim}$ package of \citet{melnykov2012mixsim}.
Here, the package allows for control of level of overlap between the
mixture components via a parameter $\bar{\omega}$ (larger implies
greater overlap), which has default value $0.05$. Here, we assess
situations where $\bar{\omega}\in\left\{ 0.05,0.5\right\} $.

We then estimate the MLE $\hat{\bm{\theta}}_{n}$ using data $\mathcal{Z}_{n}$.
An additional $N=1000$ pairs $\mathcal{Z}_{N}^{\prime}=\left\{ \left(\bm{X}_{i}^{\prime},Y_{i}^{\prime}\right)\right\} _{i=1}^{N}$
is generated from the same data generating process as $\mathcal{Z}_{n}$.
Rule (\ref{eq: MAP}) is then applied to estimate each $Y_{i}^{\prime}$
via $\hat{y}\left(\bm{X}_{i}^{\prime}\right)$. The accuracy of the
prediction is then recorded as
\[
\text{ACC}=N^{-1}\sum_{i=1}^{N}\left\llbracket Y_{i}^{\prime}=\hat{y}\left(\bm{X}_{i}^{\prime}\right)\right\rrbracket \text{,}
\]
where $\left\llbracket \mathsf{A}\right\rrbracket =1$ if statement
$\mathsf{A}$ is true and $\left\llbracket \mathsf{A}\right\rrbracket =0$,
otherwise. We also compute the accuracy of predicting the responses
of $\mathcal{Z}_{N}^{\prime}$ via the covariates, using the logistic
regression rule and SVM rule (\ref{eq: SVM rule}), where the respective
models are estimated using the data $\mathcal{Z}_{n}$, only. Here,
logistic regression and SVM are implemented using the $\mathtt{glm}$
function and the $\mathtt{svm}$ function (in the package $\mathsf{e1071}$;
\citealp{Meyer:2019aa}) in $\mathsf{R}$, respectively. 

\begin{figure}
\begin{centering}
\includegraphics[width=10cm]{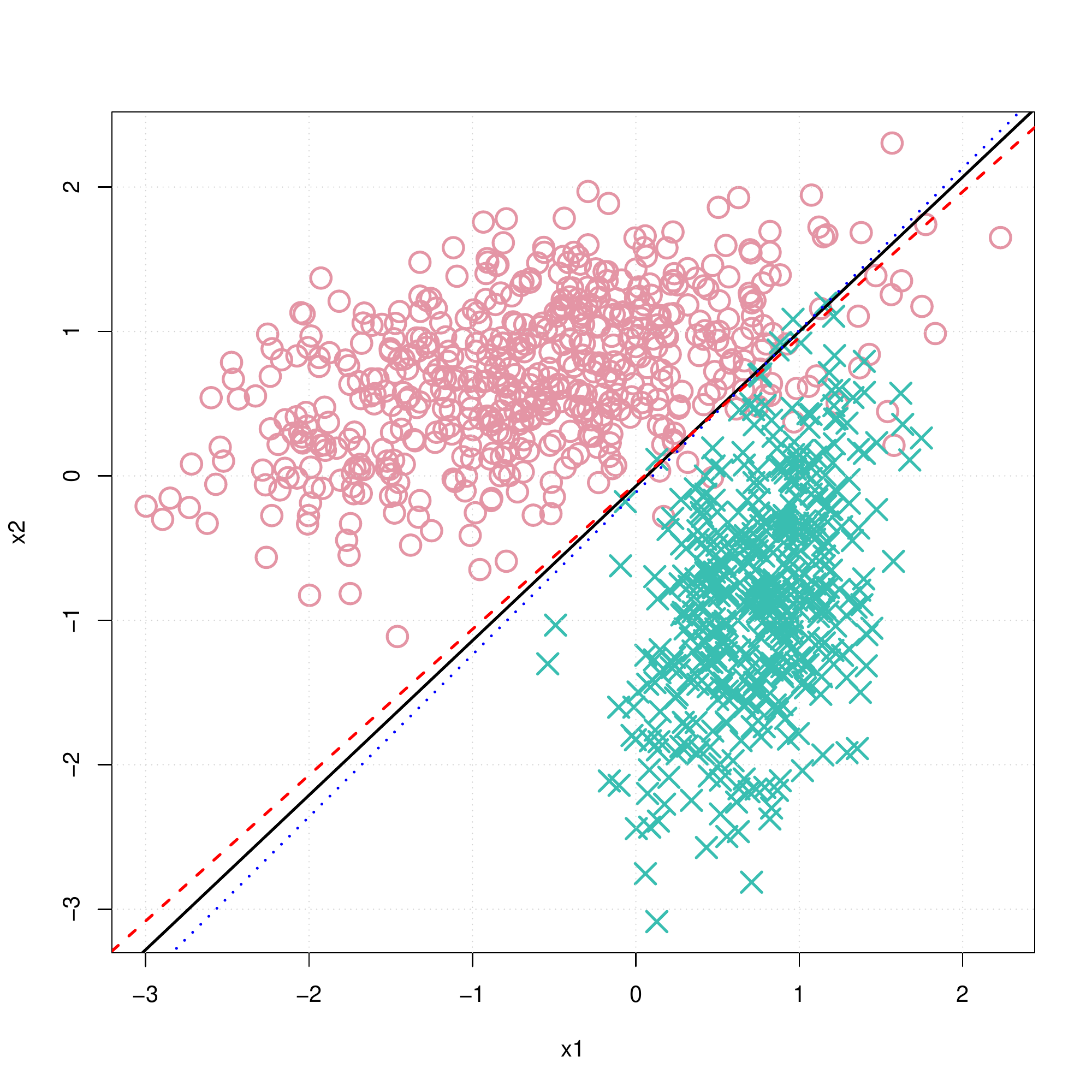}
\par\end{centering}
\centering{}\caption{\label{fig: Discrim}Scatter plot of the data $\mathcal{Z}_{N}^{\prime}$
for an instance of the $\left(d,n,\bar{\omega}\right)=\left(2,1000,0.05\right)$
simulation scenario. Here, circles and crosses indicate that $y_{i}$
equals $-1$ or $1$, respectively. The solid, dashed, and dotted
lines represent the decision rule (\ref{eq: MAP}), and the logistic
regression and SVM rules, respectively.}
\end{figure}

The experiment is repeated $R=100$ times for each combination of
$\left(d,n,\bar{\omega}\right)$. The accuracies for each of the three
assessed methods are averaged and a standard deviation is computed.
These results are presented in Table\ref{tab:Accuracy}. Example decision
boundaries for each of the three prediction rules for the $\left(d,n,\bar{\omega}\right)=\left(2,1000,0.05\right)$
case are visualized in Figure \ref{fig: Discrim}.

\begin{table}
\caption{\label{tab:Accuracy}Accuracies averaged over 100 replications (along
with standard deviations, in italic) of predictions using Rule (\ref{eq: MAP}),
logistic regression (LR), and SVM are provided for various combinations
of dimension $d$, sample size $n$, and separation coefficient $\bar{\omega}$.}

\centering{}%
\begin{tabular}{|cc|ccc|ccc|}
\hline 
 &  &  & $\bar{\omega}=0.05$ &  &  & $\bar{\omega}=0.5$ & \tabularnewline
$n=$ & $d=$ & (\ref{eq: MAP}) & LR & SVM & (\ref{eq: MAP}) & LR & SVM\tabularnewline
\hline 
100 & 2 & 0.963 & 0.963 & 0.964 & 0.663 & 0.662 & 0.670\tabularnewline
 &  & \emph{0.016} & \emph{0.016} & \emph{0.018} & \emph{0.074} & \emph{0.073} & \emph{0.070}\tabularnewline
 & 5 & 0.952 & 0.950 & 0.959 & 0.600 & 0.599 & 0.604\tabularnewline
 &  & \emph{0.014} & \emph{0.015} & \emph{0.009} & \emph{0.051} & \emph{0.052} & \emph{0.047}\tabularnewline
\hline 
1000 & 2 & 0.967 & 0.968 & 0.967 & 0.691 & 0.689 & 0.702\tabularnewline
 &  & \emph{0.014} & \emph{0.014} & \emph{0.015} & \emph{0.059} & \emph{0.059} & \emph{0.049}\tabularnewline
 & 5 & 0.966 & 0.966 & 0.966 & 0.614 & 0.614 & 0.620\tabularnewline
 &  & \emph{0.007} & \emph{0.007} & \emph{0.007} & \emph{0.049} & \emph{0.049} & \emph{0.047}\tabularnewline
\hline 
\end{tabular}
\end{table}

We now discuss some observations regarding Table \ref{tab:Accuracy}.
Firstly, all three methods appear to perform equally well across each
of the simulation scenarios. However, there is a tendency for SVM
to perform better than Rule (\ref{eq: MAP}), which also has a tendency
of performing equal or better than logistic regression. This ordering
makes some sense as Rule (\ref{eq: MAP}) is a probabilistic version
of the usual SVM rule (\ref{eq: SVM rule}), and it is also constructed
in a manner similar to that of logistic regression. We note that both
our model and logistic regression have the advantage over SVM in that
they both generate posterior probabilities of $Y^{\prime}$ given
a fixed value of $\bm{X}^{\prime}$, whereas SVM does not since it
is not probabilistic in construction. Thus, the posterior probabilities
of can be calculated directly using our method and logistic regression,
whereas SVM requires an approximate calculation via techniques such
as those of \citet{platt1999probabilistic} and \citet{lin2007note}.

Next, we observe that our usual intuition regarding difficulty of
prediction is met by these results. That is, as $n$ increases, accuracy
improves, since more data is used to learn the prediction models.
Further, greater dimensionality $d$ decreases accuracy for each fixed
$n$ and $\bar{\omega}$, since the greater dimensionality increases
the model complexity of the model. Lastly, increasing overlap drastically
decreases prediction accuracy, since the heterogeneity of the data
becomes more difficult to recognize.

\section{\label{sec:Applications}Applications}

\subsection{Wells data}

\begin{figure}
\begin{centering}
\includegraphics[width=10cm]{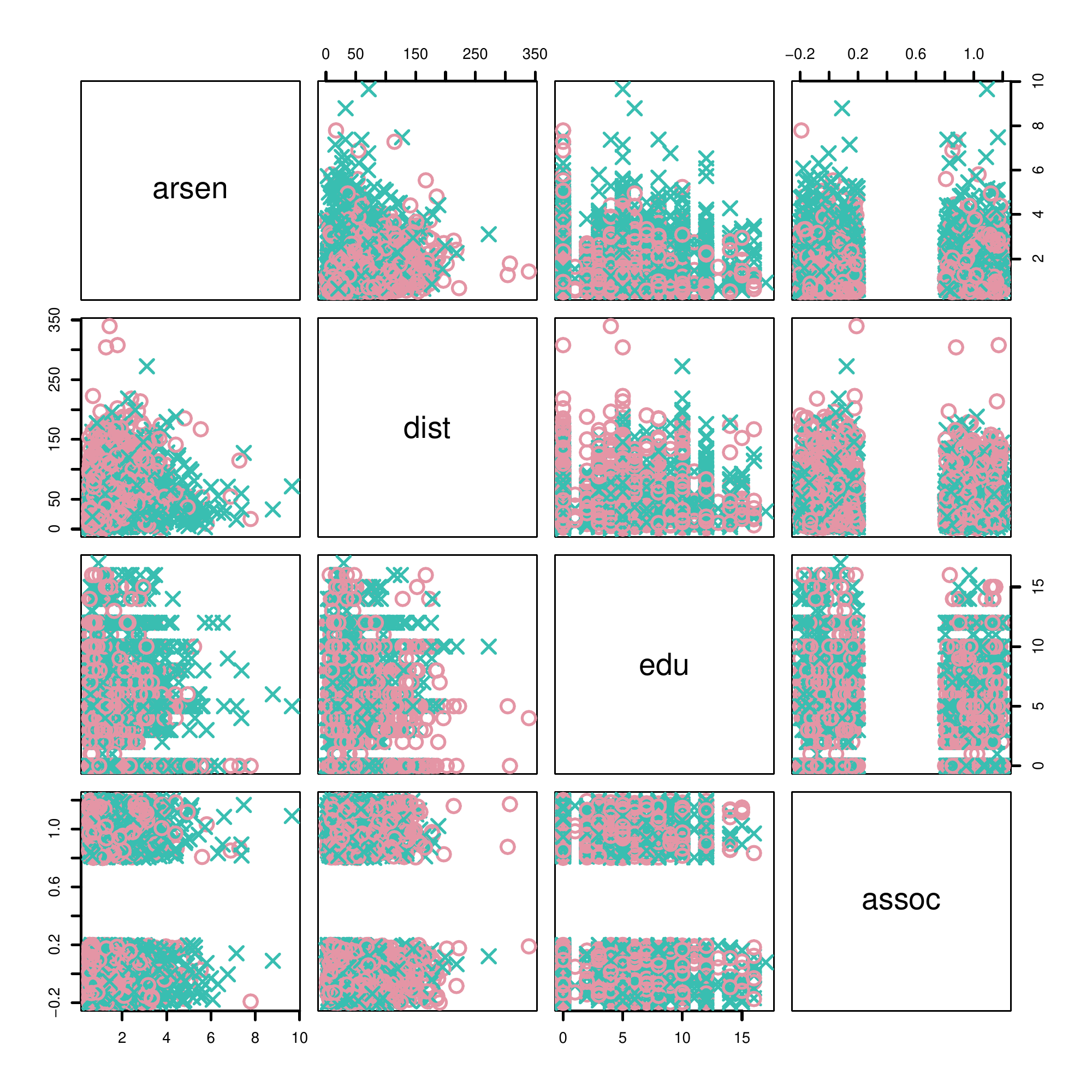}
\par\end{centering}
\centering{}\caption{\label{fig: Wells data} The pairwise scatter plots of the covariates
of the $\mathtt{Wells}$ data are plotted. Observations that correspond
to a response of $y_{i}=-1$ are plotted as circles. Observations
corresponding to $y_{i}=1$ are plotted as crosses.}
\end{figure}

We investigate the $\mathtt{wells}$ data set attributed to \citet{Gelman:2007aa},
from the $\mathsf{carData}$ package \citep{Fox:2019aa}. The data
are obtained from households in an area of Arahazar Upazila, Bangladesh,
where people were exposed to unsafe levels of arsenic in their well
water supply. The data consists of $n=3020$ households, where the
response of interest $y_{i}$ indicates whether the household $i$
switched from using their arsenic contaminated well to a safer one.
Here, $y_{i}=-1$ indicates that the household did not switch, whereas
$y_{i}=1$ indicates a switch of water supply. In order to characterize
the switching behavior, the level of arsenic contamination in the
original well of the household in hundreds of micrograms per liter
(arsen), the distance to the closest known safe well (dist), the education
level of the head of the household in years (edu), and an indicator
as to whether any members were associated with a community organization
(assoc; 1 indicates an association) were also measured as covariates
$\bm{x}_{i}$. A plot of the data appears in Figure \ref{fig: Wells data}.

To draw inference from these data, via MLE, we fit both a logistic
regression model (using $\mathtt{glm}$) and model (\ref{eq: svmit}).
The log-likelihoods of the estimated logistic regression and model
(\ref{eq: svmit}) were $-1953.91$ and $-1953.32$, respectively.
This implies that (\ref{eq: svmit}) provided a slightly better fit
to these data, but the closeness of the two log-likelihood outcomes
indicates that the inference drawn from both models should be similar.

The estimated intercept term for logistic regression was $\tilde{\alpha}_{n}=-0.1567$
$\left(0.1006\right)$, and the coefficients for each of the covariates
were estimated to be $\tilde{\beta}_{\text{arsen},n}=0.4670$ $\left(0.0452\right)$,
$\tilde{\beta}_{\text{dist},n}=-0.0090$ $\left(0.0010\right)$, $\tilde{\beta}_{\text{edu},n}=0.0424$
($0.0095$), and $\tilde{\beta}_{\text{assoc},n}=-0.1243$ ($0.0771$).
Here, the bracketed terms are asymptotic misspecification robust standard
errors, as computed via the $\mathtt{sandwich}$ function, via the
$\mathsf{sandwich}$ package \citep{Zeileis:2004aa}. Using Wald tests
for the hypotheses $\text{H}_{0}:\beta_{j}=0$ versus $\text{H}_{1}:\beta_{j}\ne0$,
we found that arsen, dist, and edu were all significant at at least
the $\alpha=10^{-5}$ level, under asymptotic normality. We found
that assoc was not significant at any $\alpha<0.1$ level.

Moving onto model (\ref{eq: svmit}), we estimated the intercept term
to be $\hat{\alpha}_{n}=-0.0871$ ($0.0505$), via MLE. The corresponding
estimates for the coefficients of the covariates were $\hat{\beta}_{\text{arsen},n}=0.2407$
$\left(0.0230\right)$, $\tilde{\beta}_{\text{dist},n}=-0.0045$ $\left(0.0005\right)$,
$\tilde{\beta}_{\text{edu},n}=0.0210$ ($0.0048$), and $\tilde{\beta}_{\text{assoc},n}=-0.0594$
($0.0387$). Here, the bracketed terms are asymptotic standard errors,
computed using Proposition \ref{prop: asymptotic norm} and expressions
(\ref{eq: est A}) and (\ref{eq: Est B}). Wald tests for the hypotheses
$\text{H}_{0}:\beta_{j}=0$ versus $\text{H}_{1}:\beta_{j}\ne0$ found
that arsen and dist were significant at the $\alpha=10^{-5}$ level,
edu was significant at the $\alpha=10^{-4}$ level, and assoc was
not significant at any $\alpha<0.1$ level. 

As expected both logistic regression and model (\ref{eq: svmit})
provided very similar inference, as we notice that all corresponding
coefficients are of the same sign. Furthermore, both models concluded
that there were significant effects due to arsen, dist, and edu, but
not due to assoc.

\subsection{Spam data}

We next investigate the $\mathtt{spam7}$ data set from the $\mathsf{DAAG}$
package of \citep{maindonald2006data}. These data contain $n=4601$
observations regarding features of emails, where the response to be
predicted is the indicator as to whether the email is spam: $y_{i}$,
which equals to $-1$ if it is not spam, and $1$ otherwise. The $d=7$
covariates stored in $\bm{x}_{i}$ by which $y_{i}$ may be conditionally
dependent upon are the total length of words in capitals, number of
occurrence of the dollar sign, number of occurrence of the bang symbol,
number of occurrences of the word 'money', number of occurrences of
the string '000', and number of occurrences of the word 'make'.

Upon fitting an SVM and model (\ref{eq: svmit}), we conduct prediction
on the data from which the models were fitted and compute the prediction
accuracies to be $0.8444$ and $0.8476$, respectively. This indicates
that model (\ref{eq: svmit}) fits the data set slightly better than
SVM. We next consider 5-fold cross-validated accuracies of the two
models (cf. \citealp{arlot2010survey}, regarding cross-validation
methods). Using the same partitioning of the data, we compute the
cross-validated accuracies to be 0.8479 (0.0095) and 0.8444 (0.0119),
respectively, where standard deviations are reported in parentheses.
We observe that both methods perform comparably in the prediction
task, although model (\ref{eq: svmit}) using Rule (\ref{eq: MAP})
yielded slightly higher accuracy levels.

\section{Concluding remarks}
\begin{rem}
A powerful concept in SVM is that of reproducing kernel Hilbert space
(RKHS) embedding. That is, instead of considering the linear map $\bm{x}^{\top}\bm{\beta}$
in (\ref{eq: SVM rule}), one considers a map $\eta:\mathbb{R}^{d}\rightarrow\mathbb{R}$,
where $\eta$ is in some RKHS $\mathcal{H}$ (cf. \citealp{steinwart2008support}).
If $\eta\left(\bm{x}\right)=\bm{\gamma}^{\top}\bm{\phi}\left(\bm{x}\right)$,
for some finite dimensional vector $\bm{\gamma}\in\mathbb{R}^{q}$
($q\in\mathbb{N}$) and map $\bm{\phi}:\mathbb{R}^{d}\rightarrow\mathbb{R}^{q}$,
then the analogous application of model (\ref{eq: svmit}) is straightforward.
That is, one simply replaces $\bm{\beta}$ with $\bm{\gamma}$ in
$\bm{\theta}$, and one replaces $\bm{x}_{i}$ by $\bm{\phi}\left(\bm{x}_{i}\right)$.
This is true for example when one considers the RKHS corresponding
to polynomial kernels of the form 
\[
\kappa\left(\bm{x},\bm{x}^{\prime}\right)=\left\langle \bm{\phi}\left(\bm{x}^{\prime}\right),\bm{\phi}\left(\bm{x}\right)\right\rangle _{\mathcal{H}}=\left(\bm{x}^{\top}\bm{x}^{\prime}+c\right)^{u}\text{,}
\]
where $c\in\mathbb{R}$ and $u\in\mathbb{N}$. Here, $\left\langle \cdot,\cdot\right\rangle _{\mathcal{H}}$
denotes the inner product of the RKHS $\mathcal{H}$. The functions
$\eta$ and $\kappa$ are related via the so-called reproducing property:
$\eta\left(\bm{x}\right)=\left\langle \eta,\kappa\left(\cdot,\bm{x}\right)\right\rangle _{\mathcal{H}}$.
In the case where $\eta$ does not correspond to some finite dimensional
mapping $\bm{\phi}$, the situation is more complicated and is beyond
the scope of this article.
\end{rem}
\begin{rem}
The coerciveness result of Proposition \ref{prop: coercivity} is
proved in terms of (\ref{eq: expected coerciveness}) in order to
facilitate the consistency result of Proposition \ref{prop: consistency}.
However, we may consider instead coerciveness of the negative log-likelihood
function $-l_{n}\left(\bm{\theta}\right)$, without any probabilistic
assumptions on the data $\mathcal{Z}_{n}$ (or assuming that $\mathcal{Z}_{n}=\left\{ \left(\bm{x}_{i}^{\top},y_{i}\right)\right\} _{i=1}^{n}$
with probability one).

Without loss of generality, we assume that $y_{i}=1$ and $y_{2}=-1$,
and we write
\begin{eqnarray*}
-l_{n}\left(\bm{\theta}\right) & = & -n^{-1}\log f\left(y_{1}|\bm{x}_{1};\bm{\theta}\right)-\log f\left(y_{2}|\bm{x}_{2};\bm{\theta}\right)\\
 &  & -n^{-1}\sum_{i=3}^{n}\log f\left(y_{i}|\bm{x}_{i};\bm{\theta}\right)\\
 & = & n^{-1}\left[\tilde{h}_{1}\left(\bm{\theta}\right)+\tilde{h}_{2}\left(\bm{\theta}\right)\right]\text{,}
\end{eqnarray*}
where
\begin{eqnarray*}
\tilde{h}_{1}\left(\bm{\theta}\right) & = & \left[1-\tilde{\bm{x}}_{i}^{\top}\bm{\theta}\right]_{+}+\left[1+\tilde{\bm{x}}_{i}^{\top}\bm{\theta}\right]_{+}\text{,}
\end{eqnarray*}
and
\begin{eqnarray*}
\tilde{h}_{2}\left(\bm{\theta}\right) & = & 2\log\left[\exp\left(-\left[1-\tilde{\bm{x}}_{i}^{\top}\bm{\theta}\right]_{+}\right)+\exp\left(-\left[1+\tilde{\bm{x}}_{i}^{\top}\bm{\theta}\right]_{+}\right)\right]\\
 &  & +\sum_{i=3}^{n}\log f\left(y_{i}|\bm{x}_{i};\bm{\theta}\right)\text{.}
\end{eqnarray*}
It is easy to see that $\tilde{h}_{2}$ is bounded from below, since
$\left[1-y_{i}\tilde{\bm{x}}_{i}^{\top}\bm{\theta}\right]_{+}$ is
bounded from below by 0 and due to the bounding of $h_{2}$ (from
Section \ref{subsec:Existence}). Thus, by the fact that the sum of
a coercive function and a function that is bounded from below is coercive,
we are only required to establish conditions under which $\tilde{h}_{1}$
is coercive. Here, we use the fact that $\tilde{h}_{1}$ is convex
and Corollary 2.5.3 of \citet{Auslender:2002aa}, which implies that
$\tilde{h}_{1}$ is coercive if the function $\tilde{h}_{1,\infty}\left(\bm{\theta}\right)>0$
for all $\bm{\theta}\in\mathbb{T}\backslash\left\{ \mathbf{0}\right\} $,
where
\begin{align*}
\tilde{h}_{1,\infty}\left(\bm{\theta}\right) & =\lim_{t\downarrow0}\,t\tilde{h}_{1}\left(t^{-1}\bm{\theta}\right)\\
 & =\frac{\left|\tilde{\bm{x}}_{1}^{\top}\bm{\theta}\right|+\left|\tilde{\bm{x}}_{2}^{\top}\bm{\theta}\right|+\tilde{\bm{x}}_{2}^{\top}\bm{\theta}-\tilde{\bm{x}}_{1}^{\top}\bm{\theta}}{2}\text{.}
\end{align*}

If we consider all the possible sign combinations of $\bm{x}_{1}^{\top}\bm{\theta}$
and $\bm{x}_{2}^{\top}\bm{\theta}$, we end up with the following
conditions that ensure $\tilde{h}_{1,\infty}\left(\bm{\theta}\right)>0$:
(1) $\tilde{\bm{x}}_{1}^{\top}\bm{\theta}\ge0$ and $\tilde{\bm{x}}_{2}^{\top}\bm{\theta}>0$,
(2) $\tilde{\bm{x}}_{1}^{\top}\bm{\theta}\le0$ and $\tilde{\bm{x}}_{2}^{\top}\bm{\theta}>0$,
(3) $\tilde{\bm{x}}_{1}^{\top}\bm{\theta}<0$ and $\tilde{\bm{x}}_{2}^{\top}\bm{\theta}\ge0$,
and (4) $\tilde{\bm{x}}_{1}^{\top}\bm{\theta}<0$ and $\tilde{\bm{x}}_{2}^{\top}\bm{\theta}\le0$.
Thus a minimal set of assumptions for the coerciveness of $-l_{n}\left(\bm{\theta}\right)$
is that $y_{1}=1$, $y_{2}=-1$, and that $\bm{x}_{1}$ and $\bm{x}_{2}$
are such that for any $\bm{\theta}\ne\mathbf{0}$, one of situations
(1)--(4) is true. This is sufficient for guaranteeing the existence
of the MLE (\ref{eq: MLE}). One situation when these conditions are
fulfilled is if $\bm{x}_{1}$ and $\bm{x}_{2}$ have no zero elements,
and if $\bm{x}_{2}=C\bm{x}_{1}$, for some $C>0$.

The conditions above can also be used to establish the existence of
a solution to problem (\ref{eq: approximate MLE}) and to the $\lambda=0$
case of the SVM problem (\ref{eq: svm optim}). To the best of our
knowledge, this is the first such set of conditions for establishing
coerciveness of the SVM problem (\ref{eq: svm optim}) when $\lambda=0$.
\end{rem}
\begin{rem}
We note that the BFGS approach that we used for optimization in Sections
\ref{sec:Implementation-and-numerical} and \ref{sec:Applications}
is by no means the only methods that can be applied to solve the MLE
problem (\ref{eq: MLE}). Recently, there has been rapid development
in the research of algorithms that are provably convergent for broad
classes of non-differentiable and non-convex optimization problems.
For example, the piecewise differentiable approximation approach of
\citet{griewank2019relaxing} is applicable, here, as well as various
techniques presented in \citet{Bagirovetal:2020aa}, such as bundle
methods and model-based derivative free methods.
\end{rem}
\bibliographystyle{apalike2}
\bibliography{20200315_Bibliography}

\end{document}